\documentclass[conference]{IEEEtran}

\usepackage{amsmath}
\usepackage{times}
\usepackage{epsfig}
\usepackage{graphicx}
\usepackage{amsmath}
\usepackage{amssymb}
\usepackage{tabularx}
\usepackage{multirow}
\usepackage{float}
\floatstyle{plaintop}
\restylefloat{table}
\usepackage{listings}
\usepackage{caption} 
\usepackage{enumitem}
\usepackage{adjustbox}
\usepackage{array}
\usepackage{xcolor}
\usepackage{listings}

\definecolor{dkgreen}{rgb}{0,0.6,0}
\definecolor{gray}{rgb}{0.5,0.5,0.5}
\definecolor{mauve}{rgb}{0.58,0,0.82}

\graphicspath{{figures}{figures/images}{figures/screenshots}}
\DeclareGraphicsExtensions{.pdf,.jpeg,.png}

\usepackage{array}
\usepackage{booktabs}
\usepackage{latexsym}
\usepackage{rotating}

\begin{document}
	
	%%%%%%%%% TITLE
	\title{Partial Reconfiguration for Design Optimization}
	\author{
		\IEEEauthorblockN{Marie Nguyen, Nathan Serafin, and James C. Hoe}
		\IEEEauthorblockA{
			Carnegie Mellon University\\
			Pittsburgh, Pennsylvania \\}
		% \textbf{	\and
		% 	\IEEEauthorblockN{Robert Tamburo}
		% 	\IEEEauthorblockA{
		% 	Carnegie Mellon University\\
		% 	Pittsburgh, Pennsylvania}
		% 	\and
		% 	\IEEEauthorblockN{Srinivasa Narasimhan}
		% 	\IEEEauthorblockA{
		% 	Carnegie Mellon University\\
		% 	Pittsburgh, Pennsylvania}
		% 	\and
		%  	\IEEEauthorblockN{James C. Hoe}
		% 	\IEEEauthorblockA{
		% 	Carnegie Mellon University\\
		% 	Pittsburgh, Pennsylvania}}
	}

	% Place def's in this file

\def\mypar/#1{{\noindent\bf{#1.}}}

\def\mod/#1{\cite{dontuse}${\textsf{module}}_{#1}$}
\def\modmm/#1{\cite{dontuse}{\textsf{module}}_{#1}}

\def\task/#1{${\textsf{task}}_{#1}$}
\def\taskmm/#1{{\textsf{task}}_{#1}}

\def\area/#1{${\textit{A}}_{#1}$}

\def\latencyf/#1#2{${\textit{Lat}}_{#1}(#2)$}

\def\run/#1{${\textsf{R}}_{#1}$}
\def\runmm/#1{{\textsf{R}}_{#1}}
\def\op/#1{${\textsf{}{Op}}_{#1}$}
\def\app/#1{${\textsf{App}}_{#1}$}

\def\perff/#1#2{$P_{{\textsf{M}}_{#1}}(#2)$}

\def\out/#1{${\textsf{Out}}_{#1}$}
\def\inp/#1{${\textsf{Inp}}_{#1}$}
\def\time/#1{${\textit{Time}}_{#1}$}
\def\timer/#1#2{${\textit{Time}}_{#1}(#2)$}
\def\timermm/#1#2{{\textit{Time}}_{#1}(#2)}
\def\areab/#1{$Area_{\textsf{#1}}$}

\def\latency/#1#2{${\textit{Lat}}_{\textsf{#1}}(#2)$}
\def\latencymm/#1#2{{\textit{Lat}}_{\textsf{#1}}(#2)}
\def\latencya/#1{${\textit{Lat}}_#1$}
\def\latencyfmm/#1#2{{\textit{Lat}}_{#1}(#2)}

\def\throughput/#1#2{${\textit{Tput}}_{\textsf{#1}}(#2)$}
\def\throughputmm/#1#2{{\textit{Tput}}_{\textsf{#1}}(#2)}

\def\throughputf/#1#2{${\textit{Tput}}_{#1}(#2)$}
\def\throughputfmm/#1#2{{\textit{Tput}}_{#1}(#2)}
\def\throughputa/#1{${\textit{Tput}}_#1$}

\def\memBW/#1{${\textit{BW}}_{#1}$}
\def\memBWmm/#1{{\textit{BW}}_{#1}}

% ``TIGHTLIST'' ENVIRONMENT (no para space between items, small indent)
\newenvironment{tightlist}%
{\begin{list}{$\bullet$}{%
    \setlength{\topsep}{0in}
    \setlength{\partopsep}{0in}
    \setlength{\itemsep}{0in}
    \setlength{\parsep}{3pt}
    \setlength{\leftmargin}{1.5em}
    \setlength{\rightmargin}{0in}
    \setlength{\itemindent}{0in}
}
}%
{\end{list}
}

\def\justcomment/#1{\footnote{\color{red} #1}}
\def\inserting/#1#2{{\color{blue} {#1}}\footnote{\color{red} #2}}
	\maketitle
	
	% As a general rule, do not put math, special symbols or citations
	% in the abstract
	
	\begin{abstract}
	
FPGA designers have traditionally shared a similar design methodology with ASIC designers. Most notably, at design time, FPGA designers commit to a fixed allocation of logic resources to modules in a design. At runtime, some of the occupied resources could be left idle or under-utilized due to hard-to-avoid sources of inefficiencies (e.g., operation dependencies). With  partial reconfiguration (PR), FPGA resources can be re-allocated over time. Therefore, using PR, a designer can attempt to reduce idleness and under-utilization with better area-time scheduling.

In this paper, we explain when, how, and why PR-style designs can improve over the performance-area Pareto front of ASIC-style designs (without PR). We  first introduce the concept of area-time volume to explain why PR-style designs can improve upon ASIC-style designs. We identify resource under-utilization as an opportunity that can be exploited by PR-style designs. We then present a first-order analytical model to help a designer decide if a PR-style design can be beneficial. When it is the case, the model points to the most suitable PR execution strategy and provides an estimate of the improvement. The  model is validated in three case studies.

\end{abstract}
\vspace{-3pt}
	
	% For peer review papers, you can put extra information on the cover
	% page as needed:
	% \ifCLASSOPTIONpeerreview
	% \begin{center} \bfseries EDICS Category: 3-BBND \end{center}
	% \fi
	%
	% For peerreview papers, this IEEEtran command inserts a page break and
	% creates the second title. It will be ignored for other modes.
	%\IEEEpeerreviewmaketitle
	
	\section{Introduction}
\label{sec:intro}

%abstract + intro max = 3 columns

\mypar/{Motivations} Today, with growing emphasis on deploying Field Programmable Gate Arrays (FPGAs) for computing, we are starting to see FPGAs' reprogrammability being recognized as a deciding feature in selecting FPGAs over ASICs \cite{7783710}. Yet, partial reconfiguration (PR), which allows parts of an FPGA to be reconfigured at millisecond timescales, remains an under-appreciated capability.  

This paper explores the questions of when, how, and why FPGA designers should consider using PR.  The discussions in this paper focus on the use of PR in challenging design scenarios that have to deliver required performance under strict area, cost, power, and energy constraints (e.g.,~\cite{wang2019live, Kim:2019:PNP:3316781.3323484}). This work is particularly relevant to AI-driven applications at the Edge (e.g.,~\cite{video_analytics, megh, wang2019live, Kim:2019:PNP:3316781.3323484}) that (1) are deployed on low-end FPGAs due to cost, power, and size concerns, and (2) need to accelerate many compute intensive tasks with stringent latency or throughput requirements (\cite{wang2019live}, \cite{Chen:2017:ESL:3132211.3134458, 8737478}).

\mypar/{Shortcomings of ASIC-Style Designs} To accelerate these constrained applications on the FPGA, designers typically commit, at design time, to a fixed allocation of logic resources to modules. We refer to this design as an ASIC-style design. At runtime, some of the occupied resources could be left idle or under-utilized due to hard-to-avoid sources of inefficiencies (e.g., operation dependencies) which may occur even in a highly-optimized design. Under-utilization may result in (1) the design not running at the desired performance given an area budget, or (2) the design running at the desired performance but being too big to fit in the given area.

\begin{figure}[t]
	\centering
	\includegraphics[width=0.45\textwidth]{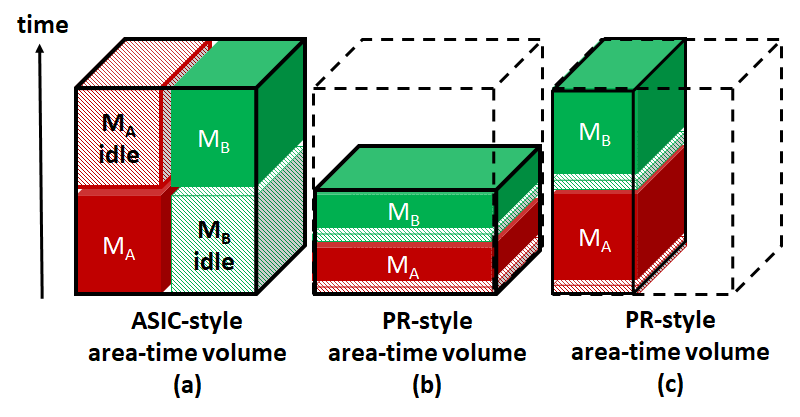}
	\caption{In an ASIC-style design, logic resources that are inactive still occupy the fabric. In a PR-style design, under-utilization can be reduced  with better area-time scheduling.   }
	\label{fig:slack}
\end{figure} 

\mypar/{PR-Style Designs to Reduce Under-Utilization}  Using PR, a designer can attempt to reduce under-utilization by changing the allocation of resources over time.  In this paper, we identify under-utilization of resources as an opportunity that can be exploited by PR-style designs to improve upon ASIC-style designs. We refer to a PR-style design as a design in which {\em{logic resources are allocated to different modules of one design over time. In return, a PR-style design may be faster and/or smaller than an ASIC-style design}} (illustration in Figure \ref{fig:slack}). 

%\noindent
\mypar/{This work: when, how and why PR} To address the questions of when, how, and why PR, this paper develops a set of PR execution strategies (allocation and scheduling) applicable to a range of non-trivial applications. An application consists of a set of tasks, and each task is accelerated by a hardware module. Modules can be dependent, execute concurrently, and have multiple implementation variants with different performance-area trade-offs. Dependent modules share data either through (1) external memory or (2) on-chip memory. The paper proposes a first-order analytical model to help a designer (1) determine a suitable PR execution strategy and (2) analyze the throughput and latency of ASIC-style and PR-style designs. The model enables quick exploration of the design space to help decide if a PR-style design can be beneficial for a given problem. The effectiveness of this model is examined in three compute-bound case studies involving computer vision and machine learning tasks. 

The contributions of this paper are:

\begin{tightlist}
	%\item developing the concept of area-time volume and identifying resource under-utilization as a condition for PR-style designs to potentially improve over ASIC-style designs 
	\item developing a set of PR execution strategies for practical design scenarios
	\item developing a first-order performance model to estimate ASIC-style and PR-style designs' performance  
	\item demonstrating the effectiveness of the performance model with three case studies of implemented designs.
\end{tightlist}

%\mypar/{Outline}  Following this introduction, Section~\ref{sec:concepts} reviews background and related work. Section~\ref{sec:dpr} further develops the key concepts of slack and area-time volume. Section~\ref{sec:model} explains the first-order analytical model.  Section~\ref{sec:setup} presents the case studies and the evaluation setup. Section~\ref{sec:results} discusses the results.  Finally, Section~\ref{sec:conclusion} concludes.

	\section{Background and Related Work}
\label{sec:concepts}

\mypar/{Partial Reconfiguration} When using PR, the FPGA fabric is divided into a non-reconfigurable region (containing the I/O infrastructure) and PR regions that can be reprogrammed individually at runtime. 
Each PR region can be reprogrammed at runtime with partial bitstreams built for this region at design time.  When loading bitstreams from on-board DRAM, the time to load a PR region (PR time) is a function of the bitstream size, e.g., approximately 453 MB/sec on an Ultrascale+ device through the processor configuration access port (PCAP). 

\mypar/{Applications of PR Today} Many academic projects have explored the potential of using PR (e.g., \cite{Majer:2007:ESM:1265130.1265134, 8416274, 6239790, 10.5555/647927.739401, 7082796, 6927494, 6386907, 10.1007/3-540-46117-5_31,  10.1145/3337801.3337816, 8457647, 6132709, 1303106, 7393288}).  Commercially, PR has been mainly used in a ``role-and-shell" approach (\cite{7783710,amazon}). A static shell design provides I/O and isolation while independent designs with different functionalities, or roles, can be loaded as required (e.g.,~\cite{7783710, amazon}).  The different role designs reuse the same logic resources over time. However, each role is still an ASIC-style design. This paper does not focus on using PR in a role-and-shell approach.  % most FPGA resources are contained in a single PR region that is enclosed by a shell region. Each role design is created to use the entire PR region alone, with no consideration for sharing resources or interacting with other role designs. This use of PR does not meet our definition of a PR-style design.   

\mypar/{PR-Style Benefits}  In a PR-style design, the designer decides how the FPGA is divided into PR regions and when/which reconfigurations are needed during the design's execution. For instance, in \cite{DBLP:conf/fpl/NguyenH18}, to accelerate a vision processing pipeline, a PR region is reconfigured every few milliseconds with different pipeline stages. 

For domain-specific applications, other prior works have exploited under-utilization in ASIC-style designs, and have shown that using PR can provide area \cite{autovision}, performance \cite{Koch:2011:FHP:1950413.1950427, Arram:2015:RRA:2684746.2689066, 5424204}, power/energy \cite{4380654, benefits}, and compilation time reduction \cite{10.1145/2847263.2847341} benefits. For instance, in adaptive \cite{region_alloc} or cloud computing applications (\cite{7396187, Chen:2014:EFC:2597917.2597929, 6861604}), multiple modes or implementation variants exist for a module but only one is needed at a time depending on the context. Instead of mapping all variants of a module in an ASIC-style design, only one variant is reprogrammed on the fabric at a time. % The work in \cite{10.1145/1328554.1328561} proposed mathematical formulations to estimate the merits of PR-style designs compared to a full reconfiguration approach, but not against ASIC-style designs. 

\mypar/{Scheduling for PR-Style} A vast body of work on FPGA OSes (\cite{222551, 4101098, 6636314, So:2008:UHR:1331331.1331338, 6927488}) and on FPGA virtualization  (\cite{8533479, 6861604, 7396187}) has focused on the theory of spatial and temporal sharing, mechanisms for task preemption, or hardware and software task scheduling \cite{1510348}. Mostly, these works share the common goal of maximizing resource utilization to improve throughput, and often assume that PR time is negligible compared to compute, and/or that tasks are independent.

Building on top of prior work, this paper introduces the concept of area-time volume to make clear why PR-style designs can be beneficial. We also give practical examples of when it is the case considering both throughput, which is the metric to optimize in many applications (e.g., video analytics \cite{video_analytics,  wang2019live}, batch jobs \cite{megh, 216801}), and latency, the metric of interest for an emerging class of Edge applications that have tolerance for 100 ms-response time, and that could benefit from FPGA acceleration (\cite{Chen:2017:ESL:3132211.3134458, 8737478}).  We also account for cases where PR time can be equal to or greater than compute time.

	\section{When and how can PR help?}

\label{sec:dpr}

In this section, we use an idealized and simplified example to develop the intuitions behind when and how PR-style designs can be faster or smaller than ASIC-style designs. The next section continues with a more complete examination.

\subsection{Simplified Execution Model}

We consider an application with two dependent tasks, \task/{A} and \task/{B}; \task/{B} can start only after \task/{A} is finished. Each task runs once per execution of the application.  The latency of the application is the sum of the two dependent tasks' latencies.  Multiple implementation module variants exist for \task/{A} and \task/{B} and are characterized by the latency function \latencyf/{i}{ }. \latencyf/{i}{a} is the latency achieved by the module variant for \task/{i} using $a$ logic resources. For a given \task/{i}, larger variants have lower latency, \latencyf/{i}{a}$<$\latencyf/{i}{b} if $a>b$. 

\subsection{ASIC-Style Design}

Consider two common design objectives: (1) minimize latency given an area budget, or (2) minimize area given a latency upper bound. For simplicity, assume \latencyf/{A}{a}=\latencyf/{B}{a} for any $a$.  In that case, to achieve optimality in either objective, the total logic resources, \area/{\text{total}}, must be equally divided between \task/{A} and \task/{B}'s modules  (\area/{A}=\area/{B}=0.5\area/{\text{total}}).  The latency of the application is 2\latencyf/{A/B}{0.5A_{\text{total}}}.  Solving either optimization scenarios repeatedly for different latency or area targets will produce a set of ASIC-style implementations that trade off latency against logic resources.  Starting from this, we ask the question: can a PR-style design improve over the Pareto front of an ASIC-style design?

\subsection{PR-Style Design}

The above scenario for the ASIC-style design is shown in Figure \ref{fig:slack}.a. In this area-time volume representation of the FPGA, the fabric area is 100\% occupied by the modules for \task/{A} and \task/{B}.  However, due to the dependency between the two modules, only one of the two modules is active at a time. In other words, the ASIC-style design has under-utilization since some resources available to the design are not active all the time. 

In contrast to an ASIC-style design where resource allocation cannot change over time, it is possible to reduce under-utilization with better area-time scheduling in a PR-style design. Therefore, a PR-style design may be able to achieve a smaller area-time volume by being faster, by using fewer resources, or both. For instance, to minimize latency given the same area budget, we can allocate the entirety of \area/{\text{total}} to a module for \task/{A} first and then to \task/{B} (Figure \ref{fig:slack}.b). By doing so, the PR-style design's latency is reduced as both modules now run faster using all of the resources available.   On the other hand, a PR-style design can maintain the same latency using half the resources by allocating  0.5\area/{\text{total}} to a module for \task/{A} first and then to \task/{B} (Figure \ref{fig:slack}.c). With under-utilization reduced, both PR-style designs fit into smaller area-time volumes than the ASIC-style design. Notice in Figure~\ref{fig:slack}.b and Figure~\ref{fig:slack}.c, a small amount of under-utilization appears when switching between modules to reflect the non-zero delay to perform PR. 

\subsection{Opportunities for Improvement by PR} In ASIC-style designs, resource under-utilization stemming from data dependencies cannot be eliminated without changing the initial algorithm or implementation. In practice, under-utilization can arise in other forms. In our simplified example, we assume that module variants exist for any amount of resources.  However, module variants for a task only exist at certain performance/resource combinations in practice.  The modules selected to fit an area budget in an ASIC-style design may not sum up perfectly to use all resources.  Further, when the modules of \task/{A} and \task/{B} are executed in a pipelined fashion to improve the throughput of many independent executions, it may not be possible to find variants with equal throughput for the two tasks; in the resulting unbalanced pipeline, a too-fast stage has to stop or slow down to wait for the other stage. A more subtle example exists when implementing a generic engine capable of accelerating different algorithms or neural networks. This generalized engine consists of a superset of features to accommodate all possibilities but only a subset of features is needed at a time (e.g., NPU \cite{8416814}, DPU \cite{dpu}).  A PR-style design could potentially remove this type of inefficiencies. 

	\section{Analytical Model}
\label{sec:model}

In this section, we present our model and discuss the additional memory requirements of a PR-style design and the impact of limited memory bandwidth on design's performance. 

\begin{figure}[t]
	\centering
	\includegraphics[width=0.4\textwidth]{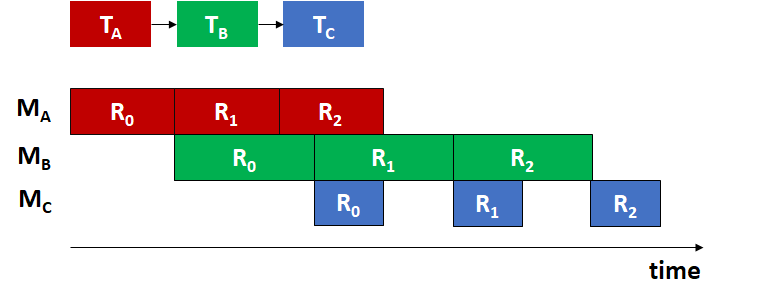}
	\caption{ Example timeline of an application with three dependent tasks accelerated by modules $M_{A}$, $M_{B}$, and $M_{C}$. } 
	\label{fig:timeline}
\end{figure} 

\subsection{Overview}

\mypar/{Optimization Goals} To derive our performance model, we consider the problem of maximizing an application's performance given an area budget. 
\begin{itemize}
	\item \textit{minimize the application's latency given an area budget $A$}. We label this problem as \textbf{min L given A}.
	\item \textit{maximize the application's throughput given an area budget $A$}. We label this problem as \textbf{max T given A}.
\end{itemize}

\mypar/{Execution Model} In this section, we consider an application with $N$ dependent tasks; each task is accelerated by a module. Dependent modules share data either through external or on-chip memory depending on data size. Though our discussion focuses on applications with dependent tasks, our model also applies if tasks are independent.  We define $I$ as the set of subscripts for tasks in the application. A single start-to-finish execution of a module is referred to as a run. If an application requires multiple independent runs, modules can execute concurrently. Figure \ref{fig:timeline} illustrates this execution model. The example application consists of three dependent tasks \task/{A}, \task/{B}, and \task/{C} accelerated by three modules. In this application, each module needs to complete three runs \run/{0}, \run/{1}, and \run/{2}. Modules execute concurrently to complete the runs as quickly as possible, subject to the dependency constraints.

We consider two performance metrics, latency and throughput. Latency is defined as the start-to-finish time required for all modules accelerating an application to complete one run (including I/O time for data read and write and compute time). Throughput is defined as the number of runs completed per unit time in steady-state.

\mypar/{Performance-Area Trade-offs} For each module, a finite set of implementation variants exists. A variant accelerating \task/{i} is characterized by its area $a_{i}$, its latency \latencyf/{i}{a_i}, and its throughput \throughputf/{i}{a_i} as functions of area. We assume that \latencya/{*} and \throughputa/{*} are monotonically increasing functions but make no further assumption on their shape, e.g., performance can scale sub-linearly or lineary with area.

\mypar/{PR-Style Design Considerations} We define $\timermm/{PR}{a}$ as the time to reconfigure a PR region of size $a$, and assume that PR time is proportional to the PR region size. 

\begin{figure}
	\centering
	\includegraphics[width=0.26\textwidth]{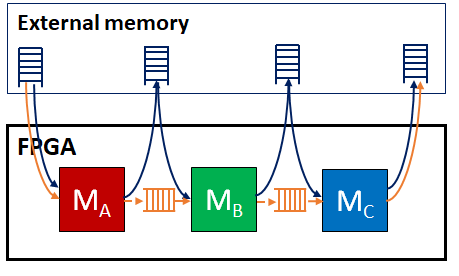}
	\caption{In an ASIC-style design, dependent modules share data through either external (blue) or on-chip memory (orange) depending on data size.}
	\label{fig:io}
\end{figure}

\subsection{ASIC-Style}
We first derive the equations for the ASIC-style design that are applicable whether dependent modules share data through external or on-chip memory (Figure \ref{fig:io}). In both cases, the number of buffers required to hold intermediate data is $N+1$.

\mypar/{Min L Given A} Let \latency/{Asic}{A} be the latency of the ASIC-style design given $A$ resources.
\begin{equation}
\latencymm/{Asic}{A} = \sum_{i \in I} \latencyfmm/{i}{a_i}, \sum_{i \in I} a_i \le A
\label{eq:latency_asic}
\end{equation}

\mypar/{Max T Given A} Let \throughput/{Asic}{A} be the throughput of the ASIC-style design given $A$ resources. 
\begin{equation}
\throughputmm/{Asic}{A} = \min(\{\throughputfmm/{i}{a_i} \mid i \in I\}), \sum_{i \in I} a_{i} \le A 
\label{eq:throughput_asic}
\end{equation}

\subsection{Ignoring PR Time: PR-Style Performance Bounds}

Ignoring PR time, we first derive the lower and upper bounds on the latency and throughput, respectively, achievable by any PR-style design presented in the next subsections.  The simplest and most efficient execution strategy is to schedule tasks serially on one PR region. Each module runs once before the PR region is reconfigured with the next module. In the best-case scenario, the PR region is of size $A$ and the highest performance variant using $A$ resources exists for all modules. 

\mypar/{Min L Given A} Let \latency/{PR,1,min}{A} be the lower bound on latency for the PR-style design with one PR region. 
\begin{equation}
\latencymm/{PR,1,min}{A} = \sum_{i \in I}  \latencyfmm/{i}{A} 
\label{eq:latency_no_pr}
\end{equation}

\mypar/{Max T Given A} Let \throughput/{PR,1,max}{A} be the upper bound on throughput for the PR-style design with one PR region.
\begin{equation}
\throughputmm/{PR,1,max}{A} = \frac{1}{\sum\limits_{i \in I} \frac{1}{\throughputfmm/{i}{A}}}
\label{eq:throughput_no_pr}
\end{equation}

\begin{figure}[t]
	\centering
	\includegraphics[width=0.38\textwidth]{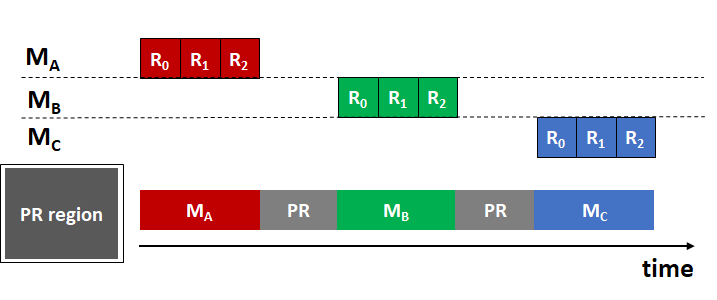}
	\caption{ Example of serialized execution in a PR-style design with one PR region when batching ($B = 3$). }   
	\label{fig:serialized_timeline}
\end{figure}

\subsection{Including PR Time: Serialized Execution on one PR Region}

When accounting for PR time and scheduling tasks serially on one PR region, each module runs once before the PR region is reconfigured with the next module. Given $N$ tasks, the PR region is reconfigured $N$ times. Compute and reconfigurations are serialized.

\mypar/{Min L Given A} Let \latency/{PR,1}{A} be the latency of the PR-style design with one PR region. 
\begin{equation}
\latencymm/{PR,1}{A} = \sum\limits_{i \in I} \latencyfmm/{i}{a_i} + N \times \timermm/{PR}{A} \\ 
\label{eq:latency_spr}
\end{equation}

Scheduling tasks serially on one PR region of the largest size may not result in the design's minimum latency.  Though using larger variants leads to a decrease in compute time, it also has the effect of increasing PR time, which may offset the speedup benefit of larger variants. In the next subsection, we discuss a scheduling alternative where compute and reconfigurations are overlapped.

\mypar/{Max T Given A: Batching to Amortize PR Time} Let \throughput/{PR,1}{A} be the steady-state throughput of the PR-style design with one PR region.
\begin{equation} 
\throughputmm/{PR,1}{A} = \frac{1}{\sum\limits_{i \in I} \frac{1}{\throughputfmm/{i}{a_i}} + N \times \timermm/{PR}{A}}
\label{eq:throughput_spr_nb}
\end{equation}

If PR time is non-trivial compared to compute time, we can amortize PR time by executing each module $B$ times (i.e. batching $B$ runs) before reconfiguring the PR region (Figure \ref{fig:serialized_timeline}). Let \throughput/{PR,1}{A} be the steady-state throughput of the PR design with one PR region when batching runs.
\begin{equation}
\throughputmm/{PR,1}{A} = \frac{B}{\sum\limits_{i \in I} \frac{B}{\throughputfmm/{i}{a_i}} + N \times {\textit{Time}}_{PR}(A)}
\label{eq:throughput_spr_b}
\end{equation}

Batching allows us to reduce the ratio of total PR time to total compute time at a greater resource cost to buffer intermediate results. Given enough buffering capacity, PR time can be almost totally amortized for large enough $B$. 

\subsection{Including PR Time: Special Cases}

%\subsection{Min L Given A: Interleaved Execution on Two PR regions}

\begin{figure}
	\centering
	\includegraphics[width=0.4\textwidth]{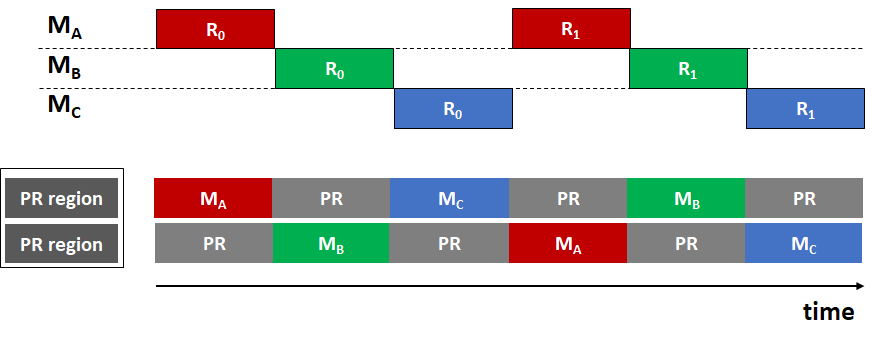}
	\caption{Interleaved execution on two PR regions. PR time can be hidden by overlapping compute and reconfiguration.}
	\label{fig:double-buffered}
\end{figure}

\mypar/{Min L Given A: Interleaved Execution on Two PR regions} When optimizing for latency, interleaving task execution on multiple PR regions allows us to overlap reconfigurations and compute to hide PR time, which may result in better latency than serializing task execution on one PR region.  Figure \ref{fig:double-buffered} shows an example of interleaved execution for $k = 2$. In this example, $\timermm/{PR}{A/2} = \latencyfmm/{i}{a_i},\forall i \in I$. By overlapping compute and reconfigurations, PR time is completely hidden. Having $k > 2$ may be beneficial provided that multiple PR regions can be reconfigured simultaneously. Simultaneous reconfiguration of multiple PR regions is not supported from a user standpoint using current FPGA tools and PR flow. In this paper, we only consider the case where $k = 2$, and define \latency/{PR,2}{A} as the latency of the PR-style design with two PR regions.
\begin{equation}
\latencymm/{PR,2}{A} = \sum\limits_{i \in I} \max( \timermm/{PR}{A/2}, \latencyfmm/{i}{a_i} )
\label{eq:latency_mpr}
\end{equation}

%\subsection{Max T Given A: Serialized Execution on $k$ PR regions} 

\mypar/{Max T Given A: Serialized Execution on $k$ PR regions} When optimizing for throughput, it is generally preferable to choose the smallest $k$ to reduce a design's complexity in terms of  buffering management since each PR region requires its own intermediate buffer. A $k$-PR region solution should be considered when appropriately large module variants are not available for all modules in a single PR region solution.

When having multiple PR regions executing in parallel (similar to $k$-way SIMD), task execution can be serialized on each PR region of size $A/k$. On each PR region, each module runs once or multiple times before the PR region is reconfigured. 
Let \throughput/{PR,1}{A/k} be the steady-state throughput of a single PR region of size $A/k$ and \throughput/{PR,k}{A} be the steady-state throughput of the PR-style design with $k$ PR regions. Assuming that $k$ reconfigurations can occur simultaneously,

\begin{equation}
\throughputmm/{PR,k}{A} = k \times \throughputmm/{PR,1}{A/k}
\label{eq:throughput_mpr}
\end{equation}

As explained previously, only one reconfiguration can happen at a time using current tools. The above throughput can still be achieved by offsetting the start of compute on each PR region by a sufficient number of PR times to ensure that two PR regions are not reconfigured simultaneously. 

\subsection{Memory Requirements in PR-style designs}

In this section, we discuss the buffering and memory bandwidth requirements of a PR-style design. Compared to an ASIC-style design, a PR-style design requires additional buffering capacity for batching and additional external memory bandwidth when faster module variants are used. A module variant is faster if it uses more resources and/or operates at a higher clock frequency. For the \textbf{Max T Given A} problem, we also model the impact of limited memory bandwidth on throughput. 
%\mypar/{Min L Given A} When optimizing for latency, modules do not execute concurrently and only run once. Therefore, the same amount of memory capacity and bandwidth is needed for ASIC-style and PR-style designs.

\mypar/{Buffering Requirement} In a PR-style design, each PR region requires two intermediate buffers to hold its intermediate input and output data. The intermediate buffers can be stored in on-chip or off-chip memory depending on the data size. The on-chip buffering option is preferred to minimize the latency and power/energy for data movement. In practice, when batching to amortize reconfiguration time, the buffering capacity required by a PR-style design exceeds the amount of on-chip memory available on current FPGAs (few MBs on large FPGAs). The amount of data to buffer can range from tens to hundreds of MBs depending on the use-case. 

If the intermediate buffers are stored in on-chip memory, additional architecture support is needed so that the output of the upstream module stored on chip is used as the input to the next module. One possible solution if to design an intermediate on-chip memory controller to connect the PR region to the intermediate buffers instead of having static, direct connections between the PR region and the buffers. The on-chip memory controller fetches the data from the appropriate intermediate buffer to send to the PR region, and writes the output from the PR region to the appropriate buffer. 

\mypar/{Max T Given A: Memory Bandwidth Requirement} When maximizing throughput given an area budget, the best strategy is to serialize module execution on one PR region. An upper bound on the memory bandwidth required by the PR-style design can be determined by considering the read and write bandwidth required by the fastest variant in the design i.e. the variant with the highest throughput. 

When the memory bandwidth required by the variant is greater than the total memory bandwidth available in the system, the variant throughput is going to be degraded by some factor proportional to the memory bandwidth required. We introduce a scaling factor \textit{F} to model the impact of limited memory bandwidth on a variant's throughput. \textit{F}  is equal to the ratio of memory bandwidth required by the variant to the memory bandwidth available in the system if the bandwidth required  by the variant is greater than the bandwidth available. Otherwise, \textit{F} is equal to 1. Let $\throughputfmm/{i, peak}{a_i}$ be the peak throughput of the module variant that accelerates \task/{i}, \memBW/{i} the bandwidth requirement of the variant, and \memBW/{total} the total bandwidth available in the system.

\begin{equation}
\resizebox{.5\textwidth}{!} 
{
$\throughputfmm/{i}{a_i} = F \times \throughputfmm/{i, peak}{a_i}, 
F=\begin{cases}
\memBWmm/{i}/\memBWmm/{total}, & \text{if \memBW/{i} $>$ \memBW/{total}} \\
1, & \text{otherwise}\\
\end{cases}
$
}
\label{eq:bandwidth_impact}
\end{equation}

%steup + results = 3 pages

\section{Experimental Setup}
\label{sec:setup}

\begin {table*}[t]
\begin{center}
	\caption [Resource utilization of the two PR-style designs $\mathbf{P_{1}}$ and $\mathbf{P_{2}}$ post place \& route] {Resource utilization of the two PR-style designs $\mathbf{P_{1}}$ and $\mathbf{P_{2}}$ post place \& route on the Ultra96 v2 board at 150 MHz. In both designs, most resources are spent for compute. In $\mathbf{P_{2}}$, the PR regions are almost equally-sized.  }
	\label{tab:pr_designs} 
	\resizebox{1\textwidth}{!}{
		\begin{tabular}{l|lll|llll}
			& &  $\mathbf{P_{1}} $& \bf{(1 PR region)}  &   & $\mathbf{P_{2}}$ & \bf{(2 PR regions)}\\ 
			\hline 
			& \bf{I/O infrastructure} & \bf{PR region}       & \bf{Total}             & \bf{I/O infrastructure}  & \bf{PR region 0}  & \bf{PR region 1} &  \bf{Total}  	 \\
			LUT  & 3366 (4.8\%)    & 61,920 (87.8\%) &  65,286 (92.5\%)     &  5231 (7.4\%)    & 28,800 (40.8\%) & 30,240 (42.9 \%)  & 64,271 (91\%)    \\  
			BRAM36Kb &    0 & 198 (91.7\%) &  198 (91.7\%) & 0 & 108 (50\%) & 108 (50\%) & 216 (100\%) \\
			DSP  &  0 &  288 (80\%) &  288 (80\%) & 0 & 144 (40\%)& 216  (60\%) & 360 (100\%) \\
			PR time (ms) & N/A & 12 & N/A & N/A & 6 & 6 & N/A  \\ 
		\end{tabular} 
	}
	
\end{center}
\end {table*}

\begin {table*}[t]
\begin{center}
	\caption[Characterization of the ASIC-style design for the activity recognition study] {Resource utilization, average memory bandwidth, and throughput of the ASIC-style design and the module variants used post place \& route on the Ultra96 v2 board at 150 MHz for the activity recognition study.  }
	\label{tab:asic_activity} 
	\resizebox{1\textwidth}{!}{
		\begin{tabular}{l|lll|lll}
			& &  \bf{Module variants} &   &  & \bf{ASIC-style} &  \\ 
			\hline 
			& \bf{hog} & \bf{cnn}       & \bf{lstm}             & \bf{I/O Infrastructure}  &\bf{Modules}  & \bf{Total} 	 \\
			
			LUT  &  15,495 (22\%)          &  14,614 (20.7\%)     &  7715 (10.9\%)    &  6082 (8.6\%)   & 37,824 (53.6\%) &   43,906 (62.2\%)   \\  
			
			BRAM36Kb & 34 (15.7\%) 		       & 92 (42.6\%) 	    & 80.5 (37.3\%)        &   0  & 206.5 (95.6\%)  &  206.5 (95.6\%)\\
			
			DSP  & 64 (17.8\%)		           &10 (2.8\%) 	    & 7 (1.9\%)   & 0 & 81 (23\%) & 81 (23\%)  \\
			Memory bandwidth (MB/s) & 23.6 & 42.7 & 3.3 & N/A & N/A & 64.6 \\
			Throughput (fps) & 30 & 16 & 271 & N/A & N/A & 16 \\ 
			%Frame latency (ms) & 33.3  & 62.5  & 3.7  & N/A  & N/A & 99.5\\ 
		\end{tabular} 
	}
	
\end{center}
\end {table*}

We develop three compute-bound applications representative of real-world applications with cost constraints \cite{cnn_lstm, flow_stereo, li2018deep}. For all studies, we use a low-end FPGA board (Ultra96 v2) with a XC7ZU3EG Zynq part that has 70,560 LUTs, 216 BRAMs and 360 DSPs. These studies serve as concrete examples of ASIC-style designs with under-utilization (due to module dependencies or modules having mismatched throughput). Each application consists of three dependent tasks, with some tasks being more compute intensive than others, which perform common vision processing such as detection or classification. Dependent modules share data through external memory since the amount of on-chip memory on the Ultra96 is not sufficient to hold the inter-module buffers in on-chip memory. Note that having more tasks per application would favor PR-style designs, since the length of the dependency chain would increase. In other words, we choose to focus on more challenging design scenarios (shorter pipelines). 

\mypar/{Design Scenario} In the studies, we solve the \textbf{max T given A} and \textbf{min L given A} problems from Section \ref{sec:model}, and also consider the problem of minimizing area given a latency upper bound, which we refer to as \textbf{given L min A}. Using our model, we search the design space to find the best-achievable ASIC-style and PR-style designs for a given problem. The best-achievable design consists of the set of module variants resulting in the design's maximum throughput, minimum latency or minimum area possible given the module variants available. We use Vivado 2019.1 to build our designs \cite{xilinx_vivado_2019}. 

\mypar/{PR-Style Designs} We consider three possible PR-style designs: (1) $\mathbf{P_{1}}$ with a single large PR region on which tasks are scheduled sequentially, (2) $\mathbf{P_{1,s}}$ with a single smaller PR region (one PR region of $\mathbf{P_{2}}$) on which tasks are scheduled sequentially, and (3) $\mathbf{P_{2}}$ with two almost equally-sized PR regions on which tasks are executed in an interleaved fashion. Table \ref{tab:pr_designs} reports the resource utilization of $\mathbf{P_{1}}$ and $\mathbf{P_{2}}$ (the PR region of $\mathbf{P_{1,s}}$ has the same size as PR region 1 of  $\mathbf{P_2}$) on the Ultra96 v2 board at 150 MHz. In both designs, most resources on the Ultra96 v2 are used for compute. The time to reconfigure a PR region through the processor configuration access port (PCAP) when partial bitstreams are stored in external DDR is 12 ms (partial bitstreams of 5.5 MB for $\mathbf{P_{1}}$) and 6 ms (partial bitstreams of 2.8 MB for $\mathbf{P_{2}}$). We use one ARM core to manage the operation of the fabric at runtime (i.e. reconfiguration of the PR regions and module execution). PR bitstreams are stored into on-board external DDR.

When optimizing for latency, we report the latency of $\mathbf{P_{1}}$, $\mathbf{P_{1,s}}$, and  $\mathbf{P_{2}}$ whenever possible. We refer to latency (or frame latency) as the time to process one input frame by the application, i.e. the time it takes for each module to run once. When optimizing for throughput, we report the throughput of $\mathbf{P_{1}}$ for different batch sizes $B$. In the context of our studies, the input to an application is a frame. When $B > 1$, the module processes $B$ frames before the PR region is reconfigured.

%\mypar/{Model Validation}  In both latency and throughput cases, we estimate the performance achieved by ASIC-style and PR-style designs using (1) our equations, (2) measured performance of modules variants, and (3) measured PR time for $\mathbf{P_{1}}$, $\mathbf{P_{1,s}}$, and $\mathbf{P_{2}}$. We report the achieved performance and percentage error to the estimated performance.

\mypar/{Performance Density} In addition to latency and throughput, we also compare the performance density of ASIC-style and PR-style designs. Performance density is defined as the number of frames processed per unit time per unit area. This metric quantifies how efficiently a design utilizes available resources. The higher the performance density, the more area-efficient the design is (less under-utilization in the area-time volume). Since there is no simple definition for area on an FPGA, we consider the resources used by the bottleneck resource as a proxy for area. For instance, if BRAM is the bottleneck as it is the case in our studies, performance density is computed as the number of frames processed per unit time per BRAM. For latency, we divide 1/latency by the number of BRAM used in the design. For throughput, we simply divide throughput by the number of BRAM used in the design.

%\begin {table*}[!t]
%\begin{center}
%	\caption [Characterization of the variants used in the activity recognition study ]{Resource utilization, throughput, and frame latency of the ASIC-style design and the module variants used post place \& route on the Ultra96 v2 board at 150 MHz for the activity recognition study.  }%The ASIC-style design has (1) unoccupied area and (2) slack since the lstm variant is four orders of magnitude faster than the hog and cnn variants. }
%	\label{tab:asic_activity} 
%	\resizebox{0.95\textwidth}{!}{
%		\begin{tabular}{l|lll|lll}
%			& &  \bf{Module variants} &   &  & \bf{ASIC-style} &  \\ 
%			\hline 
%			& \bf{hog} & \bf{cnn}       & \bf{lstm}             & \bf{I/O Infrastructure}  &\bf{Modules}  & \bf{Total} 	 \\
%			
%			LUT  &  15,495 (22\%)          &  14,614 (20.7\%)     &  7715 (10.9\%)    &  6082 (8.6\%)   & 37,824 (53.6\%) &   43,906 (62.2\%)   \\  
%			
%			BRAM36k & 34 (15.7\%) 		       & 92 (42.6\%) 	    & 80.5 (37.3\%)        &   0  & 206.5 (95.6\%)  &  206.5 (95.6\%)\\
%			
%			DSP  & 64 (17.8\%)		           &10 (2.8\%) 	    & 7 (1.9\%)   & 0 & 81 (23\%) & 81 (23\%)  \\
%			
%			Throughput (fps) & 30 & 16 & 271 & N/A & N/A & 16 \\ 
%			Frame latency (ms) & 33.3  & 62.5  & 3.7  & N/A  & N/A & 99.5\\ 
%		\end{tabular} 
%	}
%	
%\end{center}
%\end {table*}

\mypar/{Module Characterization} In the studies, we use six modules: hog \cite{hog}, cnn \cite{zqnet}, lstm \cite{Rybalkin2018FINNLLE}, viola \cite{zhou-rosetta-fpga2018}, flow \cite{flow}, and stereo (developed in-house). Each module has up to three implementation variants generated with Vivado HLS 2019.1 \cite{xilinx_hls_2019}. The variants are provided by the module developer or obtained by changing parameters in the HLS source code, such as the number of compute engines, the data precision, and the on-chip buffering size. The modules' interfaces are modified to conform to our PR region interfaces. In our studies, all PR regions have the same interfaces, namely, one AXI memory-mapped, one AXI-lite, a clock, a reset, and an interrupt. All data transfers, including data sharing between modules in the ASIC-style design, happen through external DRAM.

% In this work, we use six modules: hog \cite{hog}, cnn \cite{zqnet}, lstm \cite{Rybalkin2018FINNLLE}, viola \cite{zhou-rosetta-fpga2018}, flow \cite{flow}, and stereo (developed in-house). For each module, we generate up to three implementation variants 

Modules operate on 256$\times$256 frames, except for the lstm module which operates on 32$\times$32 frames. Modules process one frame at a time. Therefore, frame latency is the inverse of throughput, and includes both compute and data movement time. Data movement accounts for no more than 15\% of the end-to-end latency. For all variants, module throughput scales mostly linearly with its resources. The bottleneck resource for all modules is either LUTs or BRAM on the Ultra96 v2.

% We modify the interfaces of the modules so that they comply with our PR region interfaces. In our studies, all PR regions have the same interfaces, namely one AXI memory-mapped, one AXI, a clock, a reset and an interrupt port. Each module loads/stores data from/to external DDR. Dependent modules share data through external DDR.

%Figure \ref{fig:perf_vs_area} shows module variant throughput vs area (LUT, BRAM36k and DSP) after place \& route on the Ultra96 v2 board at 150 MHz.  We observe that (1) module throughput scales linearly with its resources, and (2) the bottleneck resource for all modules is either LUTs or BRAMs. 

%Another observation can be made for the lstm module (quantized). The smallest variant (7641 LUTs, 80.5 BRAM36k and 7 DSPs) operates at 87K fps, which is three orders of magnitude faster than the other modules. 
	\subsection{Model Validation: Case Study Results}
\label{sec:results_fccm2020}

In this section, we illustrate how to use our model and validate its effectiveness in three case studies. We show that (1) our first-order model allows to accurately estimate a design's throughput and latency. (2) Our analysis helps determine the most suited PR execution strategy for a problem. Notably, when optimizing for latency, it is important to evaluate both PR execution strategies (serialized execution on one PR region and interleaved execution on multiple PR regions) to find the best one for a given problem. (3) PR-style designs improve performance and performance density upon ASIC-style designs with under-utilization. (4) Given an area budget, if the ASIC-style design is too big to fit, using PR can help make the design fit and run at useful performance.  % Using the analysis, insights and equations provided in section \ref{sec:model}, the designer can quickly decide which PR strategies are best suited for a problem and estimate the relative merits of ASIC-style ad PR-style designs. 

\begin {table}[t]
\begin{center}
	\caption [Resource utilization, throughput and frame latency of variants used in $\mathbf{P_{1}}$]{Resource utilization, throughput and frame latency of the variants used in $\mathbf{P_{1}}$.  }%For each module, we use the largest variant available that fits in the PR region. }
	%Results are obtained after place \& route on the Ultra96 v2 board at 150 MHz.
	\label{tab:variants_pr1} 
	\resizebox{0.5\textwidth}{!}{
		\begin{tabular}{l|llllll}
			
			& \bf{hog} & \bf{cnn}       & \bf{lstm}  & \bf{stereo}   & \bf{flow}  & \bf{viola} 	 \\
			\hline 
			LUT  & 55,635 & 27,573  & 47,745          &  51477   &   40,509  & 42,283    \\  
			
			BRAM36Kb & 109  &180  & 144    		       &   96.5 &   195 & 91.5      \\
			
			DSP & 114 & 11 & 13    & 0  &  49 & 101   \\
			Throughput (fps) &116 & 32 & 2.1k & 240 & 180 & 41.3   \\ 
			Frame latency (ms) &8.6 & 31.2 & 0.48  &  4.2 & 5.6 & 24.2  \\ 
		\end{tabular} 
	}
	
\end{center}
\end {table}

\begin {table}[t]
\begin{center}
	\caption {Resource utilization and frame latency of the variants used in $\mathbf{P_{2}}$. }
	\label{tab:variants_pr2} 
	\resizebox{0.5\textwidth}{!}{
		\begin{tabular}{l|lllll}
			& \bf{hog} & \bf{cnn}       & \bf{lstm}  & \bf{stereo}       & \bf{flow}   	 \\
			\hline
			LUT  & 27,879 & 15,009  & 7461   & 23,551   &  20,106      \\  
			
			BRAM36Kb & 53.5  & 92 & 80.5  &  96.5 & 95.5       \\
			
			DSP & 114 & 11 & 13   &  0 &  48  \\
			
			%Throughput (fps) & 56 & 16 & 347,943  & 120 & 90   \\ 
			Frame latency (ms) & 17.9 & 62.5 & 0.87 &  8.3 &  11.1 \\ %5.6 & 24.2 &  31.2 & 0.001 \\ 
		\end{tabular} 
	}
	
\end{center}
\end {table}

%In the first case study, we explain how the analysis in section \ref{sec:model} can help the designer explore the design space more efficiently.  

% In this section, we first characterize the modules used in our studies. 

% Figure \ref{fig:perf_vs_area} show a characterization of throughput vs area (LUT, BRAM36k and DSP) of the six modules and their implementation variants post place \& route at 150 MHz on the Ultra96 v2 board. The frame latency of a module is equal to the inverse of its throughput. The frame latency of a module accounts for compute and data movement time to process a 256x256 frame. Data movement accounts for at most 15\% of the end to end latency.  We observe that (1) module throughput scales linearly with its resources and (2) the resource bottleneck when scaling module area is either LUTs or BRAM for all modules. %
%For instance, when scaling area for the stereo module, LUT is the bottleneck resource; for the CNN module, BRAM is the bottleneck. 

\mypar/{Study 1: Activity Recognition} The first case study performs activity recognition and is based on \cite{cnn_lstm}. Three dependent tasks are accelerated by a hog, a cnn and a lstm modules. This study explores the \textbf{max T given A} and \textbf{min L given A} problems. In this study, we explain how to use our model for quick design space exploration. The same methodology is used for the two other studies. 

\mypar/{Max T Given A} Table \ref{tab:asic_activity} shows the resource utilization and the throughput of the ASIC-style design and the module variants used. The ASIC-style design's throughput is equal to 16 fps and is limited by the throughput of the slowest module (cnn). The hog and lstm variants are roughly 2$\times$ and one order of magnitude faster than the cnn variant, respectively. The amount of computation per frame for the lstm variant is much less than the two other modules. Therefore, the ASIC-style design has under-utilization, and there is opportunity for PR to improve.

\begin{figure}[t]
	\centering
	\includegraphics[width=0.45\textwidth]{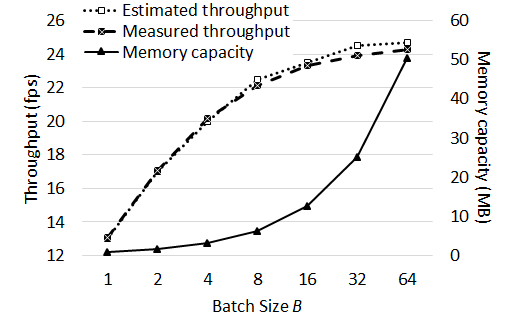}
	\caption{Throughput of $\mathbf{P_{1}}$ vs. $B$ for the first case study. }  
	\label{fig:throughput_cs1}
\end{figure}

Based on our analysis and on module variants available, batched execution on a single PR region solution ($\mathbf{P_1}$)  should provide best performance. Figure \ref{fig:throughput_cs1} shows the estimated and measured throughput, and the intermediate buffering capacity required for $\mathbf{P_1}$ vs. batch size $B$. We use equation \ref{eq:throughput_spr_b}, measured throughput variants (Table \ref{tab:variants_pr1}) and PR time (Table \ref{tab:pr_designs}) to compute these estimations.  
We observe that (1) as predicted by the model, when $B$ increases, PR time gets amortized, but with diminishing return when $B \ge 32$. (2) For all $B$, the estimated and measured throughput match within 2.35\%. (3) At $B = 64$, the throughput of the PR-style design is $24.7 \text{ fps}$, which represents a $54.4\%$ improvement over the ASIC-style design. (4) Intermediate buffering capacity linearly increases with $B$, and is equal to 50.3 MB for $B = 64$. The intermediate buffers are stored in on-board external memory (on the Ultra96, 2 GB of external DDR is available). The peak external memory bandwidth (read and write) requirement for $\mathbf{P_1}$ is 91.2 MB/s due to the hog module. This represents a 41.2\% increase over the ASIC-style design which needs on average 64.6 MB/s (Table \ref{tab:asic_activity}). 

The ASIC-style design uses $206.5$ BRAMs (95.6\% of BRAM resources) and has a performance density of $0.077$ fps per BRAM. $\mathbf{P_1}$ uses $198$ BRAMs (91.7\% of BRAM resources available) and has a performance density of $0.12$ fps per BRAM, which represents a $55.8\%$ improvement over the ASIC-style design.

%\mypar/{Summary} We validate our model and show that, for large enough $B$ ($B = 64$), the PR-style design is 54.4\% faster and improves performance density by 55.8\% compared to the ASIC-style design with slack. In this case, slack results from modules' rate mismatch, i.e. the cnn module is slower than the other modules.

% Table \ref{tab:variants_pr1} shows the resource utilization and the throughput (fps) of the hog, cnn and lstm variants used in $\mathbf{P_{1}}$. Using the throughputs in Table \ref{tab:variants_pr1} and the PR time in Table \ref{tab:pr_designs}, we estimate the throughput of $\mathbf{P_1}$ with Equation \ref{eq:throughput_spr_b}, varying $B$ from $1$ to $64$. Figure \ref{fig:throughput_cs1} shows the modeled and measured throughput. When $B=64$, the throughput of $\mathbf{P_{1}}$ is  24.7 fps. 198 BRAM resources are available in $\mathbf{P_{1}}$. The performance density of the PR-style design is 0.12 fps per BRAM.
%The PR-style design runs at 24.2 fps  which is 51.3\% faster than the ASIC-style design (16 fps). 

\begin {table*}[t]
\begin{center}
	\caption [Characterization of the ASIC-style design for the depth and motion estimation study ]{Resource utilization and latency of the ASIC-style design and module variants used post place \& route on the Ultra96 v2 board at 150 MHz for the depth and motion estimation study.} %The ASIC-style design has slack since the lstm variant is four orders of magnitude faster than the hog and cnn variants. }
	\label{tab:asic_depth} 
	\resizebox{1\textwidth}{!}{
		\begin{tabular}{l|lll|lll}
			& &  \bf{Module variants} &   &  & \bf{ASIC-style} &  \\ 
			\hline 
			& \bf{hog} & \bf{stereo}       & \bf{flow}             & \bf{I/O Infrastructure}  &\bf{Modules}  & \bf{Total} 	 \\
			
			LUT  & 27,244 (38.6\%)  &   13,767 (19.5\%)     &  10,943 (15.5\%)    &  3366 (4.8\%)   &  51,924 (73.6\%) &   55,320 (78.4\%)   \\  
			
			BRAM36Kb & 52.5 (24.3\%)  &  79.5 (36.8\%) 	    &  70.5 (32.6\%)        &   0  & 202.5 (93.8\%)  &  202.5 (93.8\%)\\
			
			DSP  &  114 (31.7\%)	& 0 	    & 44 (12.2\%)   & 0 &  158 (43.9\%) &  158 (43.9\%)  \\
			
			%Throughput (fps) & 56 & 60  & 45  & N/A & N/A & 45 \\ 
			Frame latency (ms) & 17.8  & 16.7  & 22.2  & N/A  & N/A & 56.7\\ 
		\end{tabular} 
	}
	
\end{center}
\end {table*}

\begin{figure}[t]
	\centering
	\includegraphics[width=0.5\textwidth]{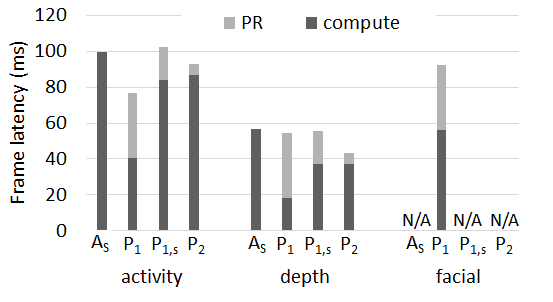}
	\caption[Frame latency of the ASIC-style design ($\mathbf{A_{s}}$) and the PR-style designs $\mathbf{P_{1}}$,  $\mathbf{P_{1,s}}$ and $\mathbf{P_{2}}$]{Frame latency of the ASIC-style design ($\mathbf{A_{s}}$) and the PR-style designs $\mathbf{P_{1}}$,  $\mathbf{P_{1,s}}$ and $\mathbf{P_{2}}$ for the three studies. }   
	\label{fig:latency}
\end{figure}

\mypar/{Min L Given A} When optimizing for latency, the ASIC-style design has under-utilization since modules are dependent (one frame processed at a time), and therefore, we expect PR to be beneficial. Figure \ref{fig:latency}.activity shows the frame latency of the latency-optimized ASIC-style design ($\mathbf{A_{s}}$), and the three PR-style designs ($\mathbf{P_{1}}$, $\mathbf{P_{1,s}}$, and $\mathbf{P_{2}}$). We estimate the latency of $\mathbf{A_{s}}$ using equation \ref{eq:latency_asic} and measured module latencies (Table \ref{tab:asic_activity}). The ASIC-style design has an estimated latency of  $99.5 \text{ ms}$, which exactly matches our measurement.

We estimate the latencies of the PR-style designs using equations \ref{eq:latency_spr} and \ref{eq:latency_mpr}, measured latencies from Tables \ref{tab:variants_pr1} and  \ref{tab:variants_pr2}, and PR time from Table \ref{tab:pr_designs}. The estimated latencies for $\mathbf{P_1}$, $\mathbf{P_{1,s}}$, and $\mathbf{P_2}$ are $76.6 \text{ ms}$, $102 \text{ ms}$, and $92.4 \text{ ms}$, respectively. The measured latencies for $\mathbf{P_1}$, $\mathbf{P_{1,s}}$, and $\mathbf{P_2}$ are $76.8 \text{ ms}$, $102.2 \text{ ms}$, and $92.6 \text{ ms}$, respectively. We observe that (1) estimated and measured latencies match within $0.26\%$, and (2) among the three PR-style designs, $\mathbf{P_{1}}$ has the smallest latency, as predicted by the model (22.8\% improvement over the ASIC-style design). Note that PR time accounts for a non-negligible fraction of the frame latency of $\mathbf{P_{1}}$ (46.9\%). However, $\mathbf{P_{1}}$ still outperforms $\mathbf{P_2}$, illustrating that the ratio of PR time to compute time should not be considered alone when optimizing for latency.

Considering performance density, $\mathbf{A_{s}}$ uses $206.5$ BRAMs and has a performance density of $0.049$ per-seconds per BRAM. $\mathbf{P_1}$ uses $198$ BRAM and has a performance density of $0.066$ per-seconds per BRAM ($34.7\%$ improvement over ASIC-style).

\mypar/{Study 2: Depth and Motion Estimation} The second case study performs depth and motion estimation, and is based on \cite{flow_stereo}. Three dependent tasks are accelerated by a hog, a stereo, and a flow module, respectively. This study explores the \textbf{min L given A} problem. 

Figure \ref{fig:latency}.depth shows the frame latency of the latency-optimized ASIC-style design ($\mathbf{A_{s}}$), and the three PR-style designs ($\mathbf{P_{1}}$, $\mathbf{P_{1,s}}$, and $\mathbf{P_{2}}$). We estimate the latency of $\mathbf{A_{s}}$ using equation  \ref{eq:latency_asic} and module latencies from Table \ref{tab:asic_depth}. The estimated latency of $\mathbf{A_{s}}$ is $56.7 \text{ ms}$ (matches the measured latency). Using the same procedure described in the first case study, we obtain latency estimations for $\mathbf{P_1}$, $\mathbf{P_{1,s}}$, and $\mathbf{P_2}$ of $54.4 \text{ ms}$, $55.3 \text{ ms}$, and $43.3 \text{ ms}$, respectively. The measured latencies for $\mathbf{A_{s}}$, $\mathbf{P_1}$, $\mathbf{P_{1,s}}$, and $\mathbf{P_2}$, are $56.7 \text{ ms}$, $54.4 \text{ ms}$, $55.3 \text{ ms}$, and $43.3 \text{ ms}$, respectively. We observe that (1) estimated and measured latencies match within 0.18\%, and (2) among all PR-style designs, $\mathbf{P_2}$ has the lowest latency, as predicted by the model ($23.6\%$ improvement over the ASIC-style design), reinforcing the fact that using the largest variants available may not achieve minimum latency.

Considering performance density, $\mathbf{A_{s}}$ uses $202.5$ BRAMs (93.8\% of BRAM resources available) and has a performance density of $0.087$ per-seconds per BRAM. $\mathbf{P_2}$ uses $216$ BRAMs, and has a performance density of $0.11$ per-seconds per BRAM ($26.4\%$ improvement over the ASIC-style design). Note that $\mathbf{P_{1,s}}$ uses only $108$ BRAMs while achieving a $2.46\%$ latency improvement compared to $\mathbf{A_{s}}$. $\mathbf{P_{1}}$ uses 2$\times$ more BRAM but only improves latency by 1.8\% compared to $\mathbf{P_{1,s}}$.  $\mathbf{P_{1,s}}$ has a performance density of $0.165$ fps per BRAM ($92.2\%$ improvement over the ASIC-style design). In a design scenario where area is to be minimized given a latency upper bound of 60 ms, $\mathbf{P_{1,s}}$ would be the best design choice.

\mypar/{Study 3: Facial Emotion Recognition} The final study performs facial emotion recognition, and is based on \cite{li2018deep}. Three dependent tasks are accelerated by a viola, a cnn and an lstm module, respectively. This study explores the \textbf{min L given A} and \textbf{given L min A} problems.

\mypar/{Min L Given A} The BRAM resources on the Ultra96 v2 are insufficient to map $\mathbf{A_{s}}$, $\mathbf{P_{1,s}}$, and $\mathbf{P_2}$. Figure \ref{fig:latency}.facial shows the frame latency of $\mathbf{P_{1}}$. Using the same procedure as in the first case study, we estimate the frame latency of $\mathbf{P_{1}}$ to be $92.2 \text{ ms}$. The measured latency is $92.1 \text{ ms}$ ($0.11\%$ error). $\mathbf{P_1}$ uses $198$ BRAMs and has a performance density of $0.055$ per-seconds per BRAM. In summary, when the ASIC-style design is too big to fit, PR can make the design fit and achieve useful performance (less than 100 ms).

\mypar/{Given L Min A} Given a latency upper bound of $100 \text{ ms}$, we want to estimate the minimum area needed by an ASIC-style design to achieve this requirement. On a larger FPGA board (Ultrascale+ 102), the ASIC-style design consisting of the smallest module variants available uses $65,987$ LUTs, $249.5$ BRAMs, and $56$ DSPs, and achieves a latency of $100.2 \text{ ms}$ post place \& route at 150 MHz. The performance density of the ASIC-style design is $0.04$ per-seconds per BRAM. Considering the PR-style design from \textbf{min L given A}, $\mathbf{P_{1}}$ improves latency by $8\%$ and performance density by $27.3\%$ compared to the ASIC-style design.

	\section{Conclusion}
\label{sec:conclusion}

This paper investigates the question of when, how and why FPGA designers should consider using PR. To address this question, we identify reducing under-utilization in ASIC-style designs as one of the main means for improvement available to PR-style designs. We then present a set of PR execution strategies to build efficient PR-style designs that can (1) be faster given an area budget or (2) smaller given a performance bound than ASIC-style designs with under-utilization. We discuss our first-order model to quickly and accurately estimate the relative merits of ASIC-style and PR-style designs in the early stage of design development.  We validate our first-order model in three study applications that serve as practical examples of ASIC-style designs with under-utilization.  Though limited, this choice of execution model and performance metrics allows us to cover a non-trivial range of design scenarios and applications (e.g., video analytics/image processing pipelines, feed-forward neural networks). 

The model relies on the existence of a module library consisting of Pareto-optimal module variants used to build the ASIC-style and PR-style designs. The accuracy of the model depends on (1) how well the library has been characterized in terms of area, latency, throughput, and memory bandwidth requirement and (2) the ability to place and route modules at the required clock frequency, which can be challenging depending on the problem. The model could be improved to account for this clock frequency uncertainty, for instance, by defining different levels of confidence based on the design's complexity. 
	\section{Acknowledgments}
\label{sec:Acknowledgments}

This work was supported in part by the CONIX Research Center, one of six centers in JUMP, a Semiconductor Research Corporation (SRC) program sponsored by DARPA. We thank Intel and Xilinx for their FPGA and tool donations.

	\bibliographystyle{ieeetr}
	\bibliography{references}
	
	\newpage

	% trigger a \newpage just before the given reference
	% number - used to balance the columns on the last page
	% adjust value as needed - may need to be readjusted if
	% the document is modified later
	%\IEEEtriggeratref{8}
	% The "triggered" command can be changed if desired:
	%\IEEEtriggercmd{\enlargethispage{-5in}}
	
	% references section
	
	% can use a bibliography generated by BibTeX as a .bbl file
	% BibTeX documentation can be easily obtained at:
	% http://mirror.ctan.org/biblio/bibtex/contrib/doc/
	% The IEEEtran BibTeX style support page is at:
	% http://www.michaelshell.org/tex/ieeetran/bibtex/
	%\bibliographystyle{IEEEtran}
	% argument is your BibTeX string definitions and bibliography database(s)
	%\bibliography{IEEEabrv,../bib/paper}
	%
	% that's all folks

\end{document}